\documentclass[12pt]{article}
\usepackage{graphicx}

\newcommand\pubnumber{}
\newcommand\pubdate{\today}

\def\taipei{Institute of Physics, Academia Sinica,
Taipei, Taiwan 115, R.O.C.}
\def\support{\footnote{Work supported in part by the National Science Council of Taiwan, R.~O.~C. under Grant Nos.~NSC-100-2112-M-001-009-MY3 and NSC-100-2628-M-008-003-MY4.}}

\def\Title#1{\begin{center} {\Large #1 } \end{center}}
\def\Author#1{\begin{center}{ \sc #1} \end{center}}
\def\Address#1{\begin{center}{ \it #1} \end{center}}

\newcommand\pubblock{\rightline{\begin{tabular}{l} \pubnumber\\
         \pubdate  \end{tabular}}}
\newenvironment{Abstract}{\begin{quotation}  }{\end{quotation}}
\newenvironment{Presented}{\begin{quotation} \begin{center}
             PRESENTED AT\end{center}\bigskip
      \begin{center}\begin{large}}{\end{large}\end{center} \end{quotation}}
\def\Acknowledgements{\bigskip  \bigskip \begin{center} \begin{large}
             \bf ACKNOWLEDGEMENTS \end{large}\end{center}}
\def\be{\begin{eqnarray}}
\def\en{\end{eqnarray}}
\def\la{\langle}
\def\ra{\rangle}

\def\CP{{\it CP}~}
\def\ol{\overline}
\def\ov{\overline}

\def\PE{{P\!E}}
\def\PA{{P\!A}}
\def\non{\nonumber}

\begin{document}
\begin{titlepage}
\pubblock

\vfill
\Title{Direct {\it CP} Violation in $D\to hh$ Decays}
\vfill
\Author{ Hai-Yang Cheng\support}
\Address{\taipei}
\vfill
\begin{Abstract}
Evidence of \CP violation in the charm sector has been observed recently by the LHCb and CDF Collaborations. The issue of whether it can be accommodated within the standard model (SM) is examined in this talk.
We conclude that the \CP asymmetry difference $\Delta a_{CP}^{\rm dir}$ between $D^0 \to K^+ K^-$ and $D^0 \to \pi^+ \pi^-$ is of order $-(0.14\sim 0.15)\%$. If the improved theoretical estimate of $\Delta a_{CP}^{\rm dir}$ in the SM remains to be a few per mille and the experimental measurement continues to be large with more statistics in the future, it will be clear evidence of physics beyond the SM in the charm sector.

\end{Abstract}
\vfill
\begin{Presented}
The 5th International Workshop on Charm Physics \\
Hawaii,  May 14--17, 2012
\end{Presented}
\vfill
\end{titlepage}
\def\thefootnote{\fnsymbol{footnote}}
\setcounter{footnote}{0}

\section{Introduction}

In the era of the luminosity frontier, there are basically two strategies for the search of New Physics (NP) in the low-energy flavor physics sector: (i) Measure those observables which are predicted to be null or almost null in the standard model (SM). Examples are {\it CP} violation in the decay $D^-(B^-)\to\pi^-\pi^0$, lepton number violation in $\tau$ decays, the CP-odd phase $\beta_s$ in $B_s\to J/\psi\phi$, like-sign dimuon asymmetry in semileptonic $B$ decays. Although the measurements of these observables will be difficult, it will be paid off once the observable of interest is detected. The advantage in this strategy is that we don't have to worry much about the SM background. (ii) Take a cue from the current anomalies observed at $B$ factories and CDF in the past years. The smoking-gun signatures of NP such as the $B$-$CP$ puzzles of $\sin 2\beta$ and $\Delta A_{K\pi}$ (the difference of \CP asymmetries in $\ov B^0\to K^-\pi^+$ and $B^-\to K^-\pi^0$) and the forward-backward asymmetry in the decay $B\to K^*\mu^+\mu^-$ are usually of 3\,$\sigma$ effects. However, many of them were already diminished by the LHCb. Nevertheless,
we did have very exciting progress in the first strategy, namely, the first evidence of \CP violation in the charm sector obtained by the LHCb collaboration \cite{LHCb} and corroborated subsequently by CDF \cite{CDF}.

A nonzero value for the difference between the time-integrated \CP asymmetries of the decays $D^0\to K^+K^-$ and $D^0\to\pi^+\pi^-$ had been reported by the LHCb Collaboration \cite{LHCb}:
\be \label{eq:LHCb}
\Delta A_{CP}\equiv A_{CP}(K^+K^-)-A_{CP}(\pi^+\pi^-)=-(0.82\pm0.21\pm0.11)\% \qquad {\rm (LHCb)}.
\en
This first evidence of \CP violation in the charm sector was later confirmed by the CDF Collaboration with the result \cite{CDF}
\be \label{eq:CDF}
\Delta A_{CP}=-(0.62\pm0.21\pm0.10)\% \qquad {\rm (CDF)} \ .
\en
The time-integrated asymmetry can be written to first order as
\be
A_{CP}(f)=a_{CP}^{\rm dir}(f)+{\la t\ra\over\tau} a_{CP}^{\rm ind} \ ,
\en
where $a_{CP}^{\rm dir}$ is the direct \CP asymmetry, $a_{CP}^{\rm ind}$ is the indirect \CP asymmetry, $\la t\ra$ is the average decay time in the sample, and $\tau$ is the $D^0$ lifetime.
The combination of the LHCb, CDF, BaBar and Belle measurements yields
$a_{CP}^{\rm ind} = -(0.025 \pm 0.231 )\%$ and $\Delta a_{CP}^{\rm dir}=
-(0.656\pm 0.154 )\%$ \cite{HFAG}. The significance of the deviation from zero is 4.3\,$\sigma$ for direct \CP asymmetry. However, for the measurements of $\Delta A_{CP}$ in (\ref{eq:LHCb}) and (\ref{eq:CDF}), neither LHCb nor CDF has measured the corresponding $A_{CP}(K^+K^-)$ and $A_{CP}(\pi^+\pi^-)$ separately. \footnote{In 2011,
the CDF Collaboration \cite{CDF1} has obtained $A_{CP}(\pi^+\pi^-)=(0.22\pm0.24\pm0.11)\%$ and $A_{CP}(K^+K^-)=-(0.24\pm0.22\pm0.09)\%$, and hence
$\Delta A_{CP}=-(0.46\pm0.31\pm0.11)\%$ based on a data sample corresponding to the integrated luminosity of $5.9$ fb$^{-1}$.}

It is commonly argued that direct \CP violation in singly Cabibbo-suppressed (SCS) $D$ decays is of order ${\cal O}([|V_{cb}^*V_{ub}|/|V^*_{cs}V_{us}|]\alpha_s/\pi)\sim 10^{-4}$ which is smaller than the LHCb measurement by two orders of magnitude. Before claiming new physics beyond the SM in the charm sector, it is crucial to have reliable SM estimate of $\Delta a_{CP}^{\rm dir}$. In this talk, we would like to address several questions: (i) Can we have a reliable theoretical estimate of strong phases and decay amplitudes? (ii) Can the LHCb measurement be accommodated within the SM? and (iii) Given the experimental results for $\Delta a_{CP}^{\rm dir}$, can one predict direct \CP violation in other SCS charm decays?

\section{Diagrammatic approach \label{sec:flavordiagram}}

In order to explore if the surprisingly large \CP asymmetries measured  by LHCb and CDF Collaborations can be explained by the SM, we need to work on a suitable theoretical framework.
In $B$ physics, there exist several QCD-inspired approaches describing the nonleptonic $B$ decays, such as QCD factorization (QCDF), pQCD and soft collinear effective theory. However, this is not the case in the $D$ sector. Until today we still don't have a satisfactory theoretical framework describing the underlying mechanism for exclusive hadronic $D$ decays based on QCD.  This is because the mass of the charm quark, being of order 1.5 GeV, is not heavy enough to allow for a sensible heavy quark expansion.  Although from time to time people have tried to apply pQCD or QCDF to hadronic charm decays, it does not make too much sense to generalize these approaches to charm decays as the $1/m_c$ power corrections are so large that the heavy quark expansion in $1/m_c$ is beyond control.

\begin{figure}[t]
\centering
\vspace*{1ex}
\includegraphics[width=2.6in]{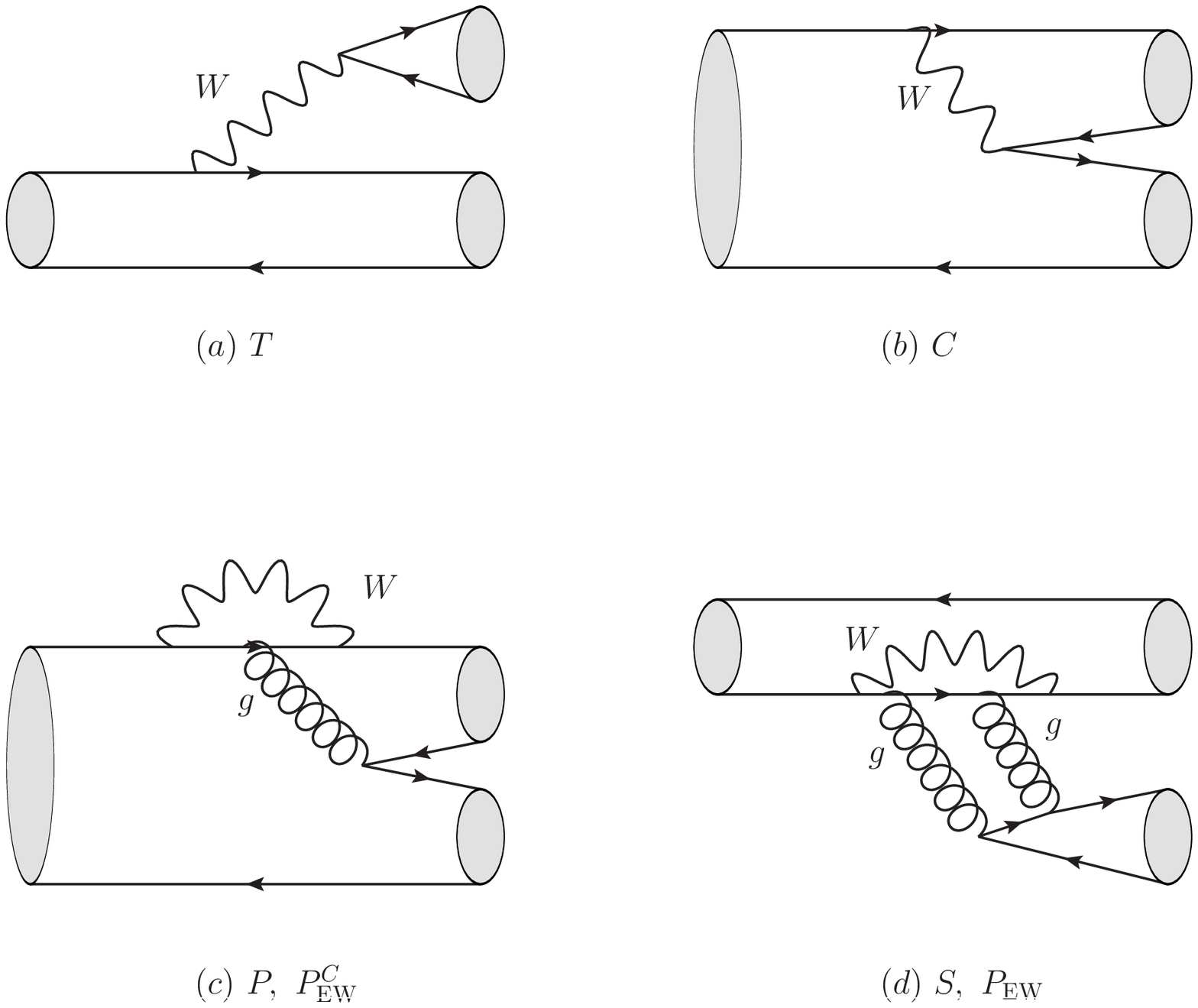}\qquad\includegraphics[width=2.6in]{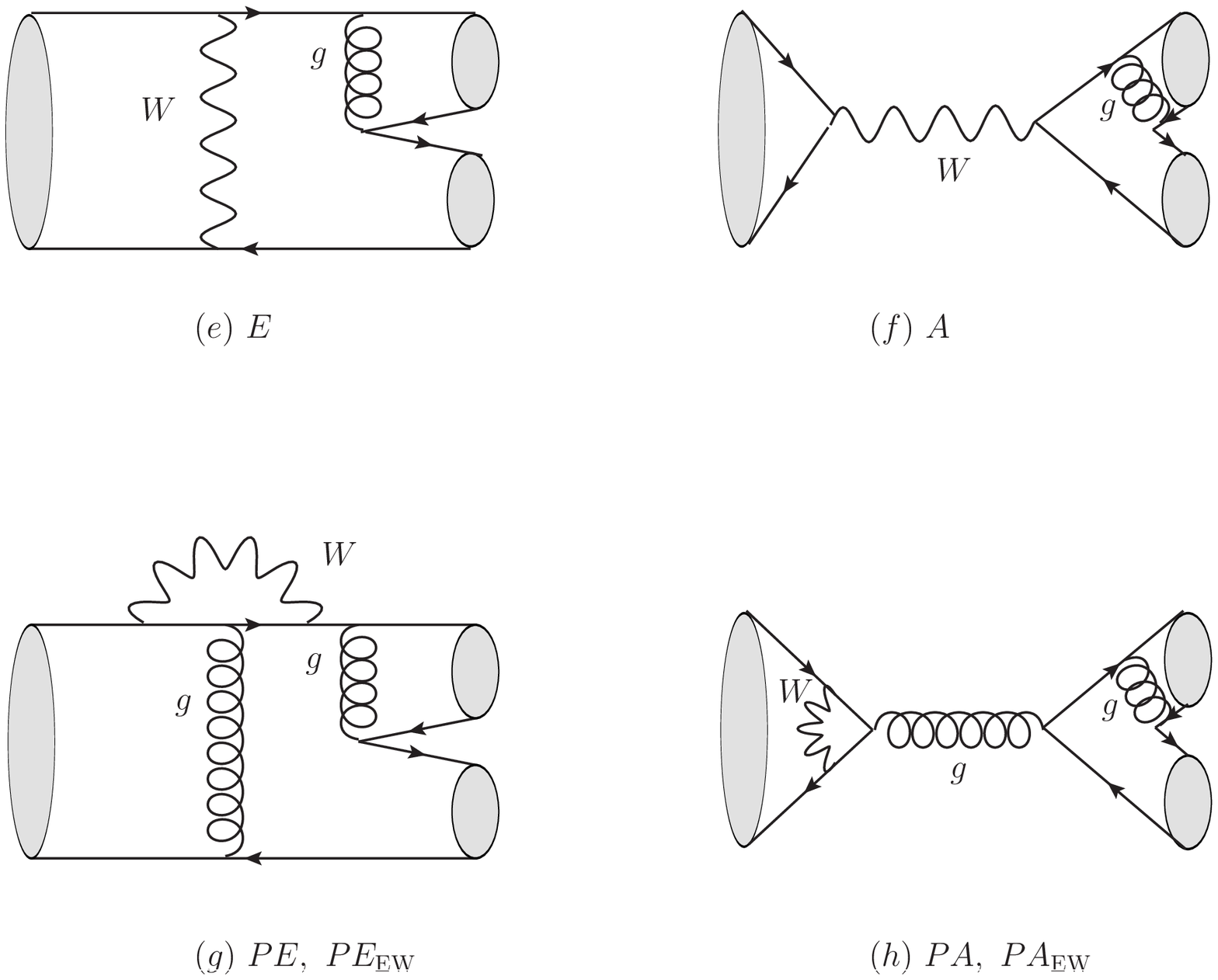}
\vspace*{-1ex}
\caption{\small Topology of possible flavor diagrams:
(a) color-allowed tree $T$, (b) color-suppressed tree $C$,
(c) QCD-penguin $P$, (d) singlet QCD-penguin $S$ diagrams with 2 (3) gluon lines for
$M_2$ being a pseudoscalar meson $P$ (a vector meson $V$),  (e) $W$-exchange $E$, (f) $W$-annihilation $A$, (g) QCD-penguin exchange $P\!E$, and
(h) QCD-penguin annihilation $P\!A$ diagrams.
The color-suppressed EW-penguin $P_{\rm EW}^C$ and color-favored EW-penguin
$P_{\rm EW}$ diagrams are obtained by replacing the gluon line from (c) and all the gluon lines from (d), respectively, by a single $Z$-boson or photon line.
The EW-penguin exchange $P\!E_{\rm EW}$ and EW-penguin annihilation $P\!A_{\rm EW}$ diagrams are obtained from (g) and (h), respectively, by replacing the left gluon line by a single $Z$-boson or photon line.
 } \label{Fig:Quarkdiagrams}
\label{Fig1}
\end{figure}

Nevertheless, we do have a powerful tool for charm physics which provides a model-independent analysis of the charmed meson decays based on symmetry, namely, the diagrammatic approach \cite{Chau,CC86,CC87}. In this approach, the topological diagrams are classified according to the topologies of weak interactions with all strong interaction effects included.  Based on flavor SU(3) symmetry, this model-independent analysis enables us to extract the topological amplitudes and it is complementary to the factorization approach. In this diagrammatic scenario, various topological diagrams are depicted in Fig.~\ref{Fig:Quarkdiagrams}.

The topological amplitudes $T,C,E,A$ can be extracted from the Cabibbo-favored (CF) $D\to PP$ decays (in units of $10^{-6}$ GeV) \cite{ChengChiang,RosnerPP08}:
\be \label{eq:PP1}
&& T=3.14\pm0.06, \qquad\qquad\qquad\quad
C=(2.61\pm0.08)\,e^{-i(152\pm1)^\circ}, \non \\
&&  E=(1.53^{+0.07}_{-0.08})\,e^{i(122\pm2)^\circ},
\qquad\quad  A=(0.39^{+0.13}_{-0.09})\,e^{i(31^{+20}_{-33})^\circ}.
\en
The topological amplitudes $C$ and $E$ extracted from the data are much larger than those expected from naive factorization. Color suppression in the $C$ amplitude and helicity suppression in $E$ are alleviated by nonfactorizable effects \footnote{Nonfactorizable contributions will render the $1/N_c$ term in the factorizable amplitude $C$ suppressed.}
and final-state rescattering. Recall that the $W$-exchange amplitude scales as $1/m_c$ and vanishes in the limit of $m_c\to\infty$. Hence, the sizable $E$ extracted from the data is clear evidence for the importance of $1/m_c$ power corrections. Considering the existence of an abundant spectrum of resonances at energies close to the mass of the charmed meson, we can even make quantitative and qualitative statements on FSI effects in weak annihilation.  As shown in Refs.~\cite{Zen,Chenga1a2}, the effect of resonance-induced FSI's can be described in a model-independent manner in
terms of the mass and width of the nearby resonances. It is found that the $E$ and $A$ amplitudes are modified by resonant FSI's as (see, {\it e.g.}, Ref.~\cite{Chenga1a2})
\be \label{eq:E,A}
E=e+(e^{2i\delta_r}-1)\left(e+{T\over 3}\right), \qquad
A=a+(a^{2i\delta_{r}}-1)\left(a+{C\over 3}\right),
\en
with
\be
e^{2i\delta_r}=1-i\,{\Gamma\over m_D-m_R+i\Gamma/2},
\en
where the short-distance contributions to the $W$-exchange amplitude $E$ and $W$-annihilation amplitude $A$ are denoted by $e$ and $a$, respectively. Therefore, even if the short-distance weak annihilation is turned off, a long-distance $W$-exchange ($W$-annihilation) contribution can still be induced from the tree amplitude $T$ ($C$) via FSI rescattering in resonance formation.

\section{Singly Cabibbo-suppressed decays}

As far as direct \CP violation is concerned, it occurs only in SCS $D$ decays.  But why? If the hadronic decay of charmed mesons proceeds only through tree diagrams, then \CP violation will not occur as only the quarks of the first two generations will get involved.
In order to have \CP nonconservation in the SM, all three-generation quarks should participate in the weak decay process. The heavy $b$ quark enters the $D$ decays through the loop diagram. Since the penguin diagram does not occur in the CF and doubly Cabibbo-suppressed (DCS) decays, direct \CP violation at tree and loop levels will manifest only in the SCS decays. Now writing the SCS decay amplitude as
\be
{\rm Amp}=\lambda_d({\rm tree}+P_d)+\lambda_s({\rm tree'}+P_s)
\en
with $\lambda_p\equiv V_{cp}^*V_{up}$, we obtain the general expression of \CP asymmetry as
\be
a_{CP}^{\rm dir}\approx {2{\rm Im}(\lambda_d\lambda_s^*)\over |\lambda_d|^2}\,\left|{D\over T}\right|\sin\delta=2\left|{\lambda_b\over \lambda_d}\right|\,\sin\gamma\left|{D\over T}\right|\sin\delta=1.2\times 10^{-3}\left|{D\over T}\right|\sin\delta,
\en
where $T$ is the dominant tree amplitude and $D$ can be a tree or penguin amplitude, $\gamma$ and $\delta$ are the weak and strong phases, respectively. For $D$ being a tree amplitude, one will expect  $10^{-3}>a_{CP}^{\rm (tree)}>10^{-5}$. For the interference between tree and penguin amplitudes, it is expected that $a_{CP}^{\rm (loop)}\sim 10^{-4}$ as $|P/T|\sim {\cal O}(\alpha_s(m_c)/\pi)$ provided that the relative strong phase is close to maximal.

\subsection{SU(3) symmetry breaking}

In order to have a reliable calculation of $a_{CP}^{\rm dir}$, an important task is to understand the SU(3) breaking effects in SCS $D\to PP$ decays.
A most noticeable example of SU(3) breaking lies in the decays $D^0\to K^+K^-$ and $D^0\to \pi^+\pi^-$. Experimentally, the rate of $D^0\to K^+K^-$ is larger than that of $D^0\to \pi^+\pi^-$ by a factor of $2.8$ \cite{PDG}, while they should be the same in the SU(3) limit. This is a long-standing puzzle since SU(3) symmetry is expected to be broken merely at the level of 30\%.

The conventional wisdom for solving the above-mentioned long-standing puzzle is that the overall apparently large SU(3) symmetry violation arises from the accumulation of several small and nominal SU(3) breaking effects in the tree amplitudes $T$ and $E$ (see {\it e.g.},\cite{Chau:SU(3)}). From the recent measurements of $\Delta A_{CP}$ by LHCb and CDF, we learn that penguin diagrams in SCS decay channels do play a crucial role for \CP violation. This leads to the conjecture that penguins may also explain the rate disparity between $D^0\to K^+K^-$ and $D^0\to \pi^+\pi^-$.

To discuss SU(3) breaking effects, we express the decay amplitudes of $D^0\to \pi^+\pi^-$ and $K^+K^-$ in terms of $\Delta P$ and $\Sigma P$:
\be \label{eq:Amp:D0pipi}
A(D^0\to \pi^+\pi^-) &=&\lambda_d(T+E+P_d+\PE_d+\PA_d)_{\pi\pi}+\lambda_s (P_s+\PE_s+\PA_s)_{\pi\pi} \non \\
&=& {1\over 2}(\lambda_d-\lambda_s)(T+E+\Delta P)_{\pi\pi}-{1\over 2}\lambda_b(T+E+\Sigma P)_{\pi\pi} \ ,
\en
where
\be \label{eq:sum&difP}
\Delta P &\equiv& (P_d+\PE_d+\PA_d)-(P_s+\PE_s+\PA_s), \non \\
\Sigma P &\equiv& (P_d+\PE_d+\PA_d)+(P_s+\PE_s+\PA_s) \ .
\en
Likewise,
\be
A(D^0\to K^+K^-)
= {1\over 2}(\lambda_s-\lambda_d)(T+E-\Delta P)_{_{K\!K}}-{1\over 2}\lambda_b(T+E+\Sigma P)_{_{K\!K}} \ .
\en

As far as the rate is concerned, we can neglect the $\lambda_b$ term. Now we consider three different cases: (i) SU(3) symmetry holds for the $T$ and $E$ amplitudes. Then a sizable  $\Delta P$ with a negative real part is needed to contribute constructively to $K^+K^-$ and destructively to $\pi^+\pi^-$; that is, $\Delta P\sim{1\over 2}Te^{-i200^\circ}$ \cite{ChengCV2}. If $U$-symmetry breaking in the amplitudes $T+E$ follows the pattern \cite{Brod}
\be \label{eq:UbreakinT}
(T+E)_{\pi\pi}=(T+E)(1+{1\over 2}\epsilon_{_T}), \qquad (T+E)_{_{K\!K}}=(T+E)(1-{1\over 2}\epsilon_{_T}),
\en
where $\epsilon_{_T}$ is a complex parameter with $|\epsilon_{_T}|\in (0,0.3)$, it has been shown in \cite{Brod} that the relation $|\Delta P/T|\sim 0.5$ still holds roughly. (ii) The realistic symmetry breaking in $T$ and $E$ amplitudes does not necessarily follow the pattern given in Eq.~(\ref{eq:UbreakinT}). Indeed, according to the factorization approach, SU(3) violation due to decay constants, meson masses and form factors leads to the robust relation $T_{_{K\!K}}/T_{\pi\pi}=1.32a_1(K\!K)/a_1(\pi\pi)\approx 1.32$\,. Assuming SU(3) symmetry again for the $E$ amplitudes, we will have $|\Delta P/T|\sim 0.15$\,.
This is also the case considered in \cite{Bhattacharya:2012}.
(iii) The large rate difference between $D^0\to K^+K^-$ and $D^0\to\pi^+\pi^-$ is entirely accounted for by SU(3) violation in the $T$ and $E$ amplitudes and $\Delta P$ is negligibly small. Owing to the observation of $D^0\to K^0\ov K^0$ through $W$-exchange and penguin annihilation diagrams and the smallness of $\Delta P$ theoretically, we have argued in \cite{ChengCV2} that the last scenario is preferred.  The $W$-exchange amplitudes can be fixed from the following four modes: $K^+K^-$, $\pi^+\pi^-$, $\pi^0\pi^0$ and $K^0\ov K^0$. Neglecting $\Delta P$ and $\lambda_b$ terms, a fit to the data yields two possible solutions
\be \label{eq:EdEs}
{\rm (I)} &&  E_d=1.19\, e^{i15.0^\circ}E, \qquad E_s=0.58\, e^{-i14.7^\circ}E
\ , \non \\
{\rm  (II)} &&  E_d=1.19\, e^{i15.0^\circ}E, \qquad E_s=1.62\, e^{-i9.8^\circ}E
\ ,
\en
where $E_q$ refers to the $W$-exchange amplitude associated with $c\bar u\to q\bar q$ ($q=d,s$).
The corresponding $\chi^2$ vanishes as these two solutions can be obtained exactly.

\section{Direct \CP violation}
\subsection{Tree-level \CP violation}

Most theory papers after the LHCb measurement focused on the decays $D^0\to K^+K^-$ and $D^0\to \pi^+\pi^-$ and emphasized the importance of interference between tree and penguin amplitudes for \CP violation. However, direct \CP asymmetry can occur at the tree level in many of the SCS decays of $D$ mesons in addition to the loop-level \CP violation \cite{Cheng1984}.
The strong point of the topological approach is that the magnitude and the relative strong phase of each individual topological tree amplitude in charm decays can be extracted from the data \cite{ChengCV1}. Hence, it becomes possible to make a reliable estimate of $a_{dir}^{\rm (tree)}$. Larger tree-level \CP asymmetries can be achieved in those decay modes with interference between $T$ and $C$ or $C$ and $E$. For example, $a_{dir}^{({\rm tree})}$ is of order $(0.7-0.8)\times 10^{-3}$ for $D^0\to \pi^0\eta$ and $D_s^+\to K^+\eta$ (see Table~\ref{tab:CPVpp}).

Direct tree-level \CP violation in $D^0\to K^0\ov K^0$ is given by
\be
a_{dir}^{({\rm tree})}(D^0\to K^0\ov K^0)=\left\{
\begin{array}{cl}
    -0.7\times 10^{-3}
      & \quad \mbox{Solution~I} \ , \\
    -1.7\times 10^{-3}
      & \quad \mbox{Solution~II} \ ,
    \end{array}\right.
\en
for the two solutions for $E_d$ and $E_s$ given in Eq.~(\ref{eq:EdEs}).
For comparison, $a_{dir}^{({\rm tree})}(D^0\to K^0\ov K^0)=1.11\times 10^{-3}$ is predicted in \cite{Li2012}.

From Table~\ref{tab:CPVpp}, we see that almost all the predicted tree-level \CP asymmetries in \cite{Li2012} are of opposite signs to ours. This is ascribed to the phase of the $W$-exchange amplitude. For CF $D\to PP$ decays, its phase is $(122\pm 2)^\circ$ with $\chi^2=0.29$ per degree of freedom [Eq.~(\ref{eq:PP1})]. For SCS decays, the phases of $E_d$ and $E_s$ [see Eq.~(\ref{eq:EdEs})] lie in the range of $107^\circ\sim 137^\circ$. Therefore, the $W$-exchange amplitude in this work is always in the second quadrant, while the $E$ amplitude in \cite{Li2012} lies in the third quadrant from a global fit to all the data of 28 CF and SCS $D\to PP$ branching fractions with $\chi^2=7.3$ per degree of freedom.  As a result, the imaginary part of $E$ in \cite{Li2012} has a sign opposite to ours, and this explains the sign difference between this work and \cite{Li2012} for $a_{dir}^{({\rm tree})}$.

\begin{table}[t]
  \footnotesize{
\begin{tabular}{l c r c r c} \hline\hline
Decay Mode &  $a_{dir}^{({\rm tree})}$(this work)~ &$a_{dir}^{({\rm tree})}$\cite{Li2012}~~~
     & $a_{dir}^{({\rm tot})}$(this work)~ & $a_{dir}^{({\rm tot})}$\cite{Li2012}~~ & Expt. \\
 \hline
$D^0\to \pi^+ \pi^-$ & $0$ & 0~~~
     & $0.95\pm0.04$ & 0.68~~  & $2.0\pm2.2$ \\
$D^0\to\pi^0 \pi^0$  & $0$ & 0~~~
     & $0.80\pm0.04$ & 0.20~~ & $1\pm48$ \\
$D^0\to \pi^0 \eta $  &  $0.82\pm0.03$ & $-0.33$~~~     & $0.08\pm0.04$ & $-0.55$~~  \\
$D^0\to \pi^0 \eta' $  &  $-0.39\pm0.02$~~ & $0.54$~~~     & $0.01\pm0.02$ & 1.99~~  \\
$D^0\to \eta\eta $ &  $-0.28\pm0.01$~~ & 0.28~~~ & $-0.58\pm0.02$~~ & 0.08~~ \\
& $-0.42\pm0.02$~~ & 0.28~~~ & $-0.74\pm0.02$~~ & 0.08~~  \\
$D^0\to \eta\eta' $  &  $0.49\pm0.02$ & $-0.30$~~~     & $0.54\pm0.02$ & $-0.98$~~   \\
 &  $0.38\pm0.02$ & $-0.30$~~~     & $0.34\pm0.02$ & $-0.98$~~   \\
$D^0\to K^+ K^{-}$   &  $0$ & 0~~~
     & $-0.42\pm0.01$~~ & $-0.50$~~ & $-2.3\pm1.7$  \\
     &  $0$ & 0~~~
     & $-0.53\pm0.02$~~ & $-0.50$~~  \\
$D^0\to K^0 \ol{K}^{0}$  & $-0.73$ & 1.11~~~ & $-0.63\pm0.01$~~ & $1.37$~~  \\
& $-1.73$ & 1.11~~~ & $-1.81\pm0.01$~~ & $1.37$~~  \\
$D^+\to \pi^+ \pi^0$   & $0$ & 0~~~
     & $0$ & 0~~ & $29\pm29$  \\
$D^+\to \pi^+ \eta $  & $0.35\pm0.06$ & $-0.54$~~~ &  $-0.74\pm0.06$~~ & $-0.52$~~  &$17.4\pm11.5$ \\
$D^+\to\pi^+ \eta' $  &  $-0.21\pm0.04$~~ & 0.39~~~    & $0.33\pm0.07$ & 1.52~~  & $-1.2\pm11.3$  \\
$D^+\to K^+ \ol{K}^{0}$~~~  &  $-0.07\pm0.06$~~ & $-0.14$~~~   & $-0.39\pm0.04$~~ & $-1.00$~~  & $-1.0\pm5.9$ \\
$D_s^+\to \pi^+ K^{0}$  &  $0.07\pm0.06$ & $0.14$~~~     & $0.45\pm0.03$ &  $1.00$~~ & $66\pm24$ \\
$D_s^+\to \pi^0 K^{+}$  & $0.01 \pm 0.11$ & 0.33~~~
     & $0.94 \pm 0.10$ & 0.72~~ & $266\pm228$    \\
$D_s^+\to K^{+}\eta$   &  $-0.71\pm0.05$~~ & $-0.19$~~~    & $-0.61\pm0.05$~~ & 0.83~~ & $93\pm152$   \\
$D_s^+\to K^{+}\eta' $  &  $0.36\pm0.04$ & $-0.41$~~~     & $-0.28\pm0.12$~~ & $-1.78$~~  & $60\pm189$   \\
\hline\hline
\end{tabular}
\caption{\small Direct \CP asymmetries (in units of $10^{-3}$) of $D\to PP$ decays. The first (second) entry in $D^0\to \eta\eta$, $\eta\eta'$, $K^+K^-$ and $K^0\ov K^0$ is for Solution I (II) of $E_d$ and $E_s$ [Eq.~(\ref{eq:EdEs})].
For QCD-penguin exchange $\PE$, we assume that it is similar to the topological $E$ amplitude [see Eq.~(\ref{eq:PE})].  World averages of experimental measurements are taken from Ref.~\cite{HFAG}. For comparison, the predicted results of $a_{dir}^{({\rm tree})}$ and $a_{dir}^{({\rm tot})}$ in \cite{Li2012} are also presented.
  \label{tab:CPVpp}}
}
\end{table}
%

\begin{table}[t]
\begin{center}
\begin{tabular}{l r r r r } \hline\hline
Decay Mode & $a_{dir}^{({\rm tree})}$
     & $a_{dir}^{({\rm t+p})}$ & $a_{dir}^{({\rm t+pa})}$ & $a_{dir}^{({\rm tot})}$  \\
 \hline
 $D^0\to \pi^+\rho^-$ & $0$ & $0.08$ & $-0.60$ & $-0.52$ \\
 $D^0\to \pi^-\rho^+$ & $0$ & $-0.05$ & $-0.22$ & $-0.28$ \\
 $D^0\to \pi^0\rho^0$ & 0 & $-0.02$ & $-0.74$ & $-0.76$ \\
 $D^0\to K^+K^{*-}$ & 0 & $-0.08$ & 0.60 & $0.52$ \\
 $D^0\to K^-K^{*+}$ & 0 & 0.07 & $0.22$ & $0.29$ \\
 $D^0\to K^0\overline K^{*0}$ & $0.73$& 0.73 & 0.73 & $0.73$ \\
 $D^0\to \overline K^0 K^{*0}$ & $-0.73$ & $-0.73$ & $-0.73$ & $-0.73$ \\
 $D^0\to \pi^0\omega$ & 0 &  $-0.01$ &  $0.53$ & $0.52$ \\
 $D^0\to \pi^0\phi$ & $0$ & 0 & 0 & 0 \\
 $D^0\to \eta\omega$ & 0.19 & 0.19 & $0.50$ & $0.50$ \\
 $D^0\to \eta'\omega$ & $-1.07$ & $-1.05$ &  $-0.91$ & $-0.89$ \\
 $D^0\to \eta\phi$ & 0 & 0 & 0 & 0 \\
 $D^0\to \eta\rho^0$ & $-0.53$ & $-0.55$ & $-0.22$ & $-0.24$  \\
 $D^0\to \eta'\rho^0$ & $0.59$ & $0.59$ & $0.21$ & $0.21$  \\
 \hline\hline
\end{tabular}
\caption{\small Same as Table \ref{tab:CPVpp} except for $D\to PV$ decays where the superscript ${\rm (t+p)}$ denotes tree plus QCD penguin amplitudes, ${\rm (t+pa)}$ for tree plus weak penguin annihilation ($P\!E$ and $P\!A$) amplitudes and ``tot'' for the total amplitude.  Due to the lack of information on the topological amplitudes $A_P$ and $A_V$, no prediction is attempted for $D^+\to PV$ and $D_s^+\to PV$ decays.
  \label{tab:CPVvp}}
\end{center}
\end{table}
%

\subsection{Penguin-induced \CP violation}

Direct \CP violation does not occur at the tree level in $D^0\to K^+K^-$ and $D^0\to\pi^+\pi^-$ decays. \CP asymmetry in these two modes arises from the interference between tree and penguin amplitudes denoted by $a_{dir}^{\rm (t+p)}$. Specifically, we have
\be \label{eq:pipiacp}
a_{dir}^{({\rm t+p})}(\pi^+\pi^-)
&\approx& 1.2\times 10^{-3} \left| {P_s+P\!E_s+P\!A_s\over T+E+\Delta P}\right|_{\pi\pi}\sin\delta_{\pi\pi} \ , \non \\
a_{dir}^{({\rm t+p})}(K^+K^-)
&\approx& -1.2\times 10^{-3} \left| {P_d+P\!E_d+P\!A_d \over T+E-\Delta P}\right|_{_{K\!K}}\sin\delta_{_{K\!K}} \ ,
\en
where $\delta_{\pi\pi}$ is the strong phase of $(P_s+P\!E_s+P\!A_s)_{\pi\pi}$ relative to $(T+E+\Delta P)_{\pi\pi}$ and likewise for $\delta_{_{K\!K}}$. Therefore, in SU(3) limit we have the relation
$a_{dir}^{({\rm t+p})}(K^+K^-)=-a_{dir}^{({\rm t+p})}(\pi^+\pi^-)$.

The QCD penguin contributions can be estimated in QCD-inspired approaches such as QCD factorization (QCDF) and pQCD or in the factorization approach. In QCDF we find
\be \label{eq:Pover T}
&& \left({P_s\over T}\right)_{\pi\pi}=0.24\, e^{-i154^\circ}, \qquad
\left({P_d-P_s\over T}\right)_{\pi\pi}=0.010\, e^{-i35^\circ}, \non \\
&& \left({P_d\over T}\right)_{_{K\!K}}=0.24\, e^{-i152^\circ}, \qquad
\left({P_d-P_s\over T}\right)_{_{K\!K}}=0.009\, e^{-i35^\circ}.
\en
The difference in the $d$- and $s$-loop penguin contractions turns out to be very small compared to the tree amplitude. It follows that
\be
\left({P_s\over T+E}\right)_{\pi\pi}=0.35\, e^{i170^\circ}, \qquad
\left({P_d\over T+E}\right)_{_{K\!K}}=0.24\, e^{i170^\circ}.
\en
Hence, $\delta_{\pi\pi}\approx \delta_{K\!K}=170^\circ$. This leads to
$a_{dir}^{\rm (t+p)}(\pi^+\pi^-)=6.7\times 10^{-5}$ and $a_{dir}^{\rm (t+p)}(K^+K^-)=-4.9\times 10^{-5}$. Therefore, QCD-penguin induced \CP asymmetries in $D^0\to \pi^+\pi^-,~ K^+K^-$ are small mainly due to the almost trivial strong phases $\delta_{\pi\pi}$ and $\delta_{K\!K}$.

\begin{figure}[t]
\centering
\includegraphics[width=0.40\textwidth]{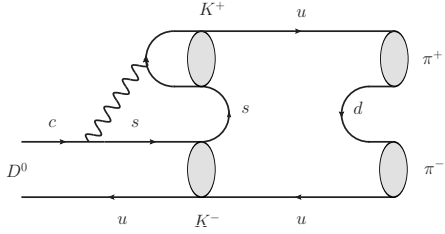}
\vspace{0.0cm}
\caption{Long-distance contribution to $D^0\to \pi^+\pi^-$ through a resonant-like final-state rescattering. It has the same topology as the QCD-penguin exchange topological diagram $P\!E$.
} \label{fig:FSI}
\end{figure}

For power corrections to QCD penguins, we shall consider QCD-penguin exchange $\PE$ and QCD-penguin annihilation $\PA$.
At the short-distance level, weak penguin annihilation contributions are found to be smaller than QCD penguin with the hierarchy $|P|>|\PE|>|\PA|$. For example, $(\PE/T)_{\pi\pi}\sim 0.04$ and $(\PA/T)_{\pi\pi}\sim -0.02$. As for long-distance contributions to weak penguin annihilation, it was pointed out in \cite{ChengCV1} that a SCS decay, for example, $D^0\to \pi^+\pi^-$, could proceed through the weak decay $D^0\to K^+K^-$ followed by a resonant-like final-state rescattering as depicted in Fig.~\ref{fig:FSI}. It has the same topology as the QCD-penguin exchange topological graph $P\!E$.  Just as the weak annihilation topologies $E$ and $A$, it is expected that weak penguin annihilation will receive sizable long-distance contributions from FSI's as well.  Hence, we shall assume that $P\!E$, $P\!E_P$ and $P\!E_V$ are of the same order of magnitude as $E$, $E_P$ and $E_V$, respectively; namely,
\be \label{eq:PE}
P\!E\approx E, \qquad P\!E_P\approx E_P, \qquad P\!E_V\approx E_V.
\en

The calculated direct \CP asymmetries for various SCS $D\to PP$ and $D\to PV$ decays are summarized in Tables~\ref{tab:CPVpp} and \ref{tab:CPVvp}, respectively. We conclude that the direct \CP asymmetry difference between $D^0 \to K^+ K^-$ and $D^0 \to \pi^+ \pi^-$ is about $-(0.139\pm 0.004)\%$ and $-(0.151\pm 0.004)\%$ for the two solutions of $W$-exchange amplitudes, respectively. A similar prediction of $-0.118\%$ was also obtained in \cite{Li2012}. Since in the SM, $\Delta a_{CP}^{\rm dir}$ arises mainly from weak penguin annihilation, we can vary the amplitude $P\!E$ to see how much enhancement we can gain. Even with the maximal magnitude $|P\!E|\sim T$ and a maximal strong phase relative to $T$, we will get $\Delta a_{CP}^{\rm dir}=-0.27\%$ which is still more than $2\sigma$ away from the current world average.

\subsection{Extraction of penguin amplitudes from the diagrammatic approach?}

The cautious reader may notice that while we have relied on the topological tree amplitudes extracted from the data to perform the analysis, we apply the theoretical model to compute the penguin amplitudes. The question is then can we extract penguin amplitudes from SCS decays within the framework of the diagrammatic approach? As shown in \cite{Bhattacharya:2012}, in principle the penguin amplitudes $\sum_{p=d,s}\lambda_p(P_p+\PE_p+\PA_p)$ can be extracted from $D^0\to \pi^+\pi^-,\pi^0\pi^0,K^+K^-$ decays, while $\sum_{p=d,s}\lambda_p(P_p+\PE_p)$ can be inferred from $D^+\to K^+\ov K^0$ and $D_s^+\to \pi^+K^0,\pi^0K^+$ provided that the tree amplitudes are those given in Eq.~(\ref{eq:PP1}). Hence, $\sum_{p=d,s}\lambda_p \PA_p$ can be determined. However, unlike topological tree amplitudes, it will be difficult to extract the topological penguin amplitudes reliably from the data. This is because the use of the topological approach relies heavily on SU(3) symmetry which leads to negligible penguin amplitudes in $D$ decays. Consequently, the extraction of penguin amplitudes depends on SU(3) breaking effects in tree amplitudes. Indeed, we have shown in Sec.~3.1 that the difference in penguin contractions $\Delta P$ is sensitive to how SU(3) symmetry breaking is treated in the tree amplitudes $T$ and $E$ . Therefore, we shall rely on theory for the estimation of the penguin contribution.

\section{Attempts for the SM interpretation}

Even long before the LHCb experiment, it had been advocated in \cite{Golden:1989qx} that the observation of \CP violation in $D$ decays would not necessarily be a signal of NP. If certain hadronic matrix elements are enhanced, in analogy to the $\Delta I=1/2$ rule of kaon decays, then \CP violation will be observable in strangeness conserving decays. This has motivated many authors to conjecture that the large direct \CP asymmetry difference $\Delta a_{CP}^{\rm dir}$ observed by LHCb and CDF could be explained or marginally accommodated in the SM \cite{Brod,Bhattacharya:2012,Kagan,PU,Feldmann,Franco}. Most of them argued that the penguin matrix elements were enhanced by some nonperturbative effects or unforeseen QCD effects.  The idea is that the QCD penguin amplitude is of $\Delta I=1/2$ transition and it will be enhanced relative to the $\Delta I=3/2$ amplitude through the $\Delta I=1/2$ rule enhancement. However, we shall show below that the enhancement of the $A_0$ amplitude relative to $A_2$ in charm decays arise predominately from the tree amplitudes.

Combining the amplitudes
\be
A(D^0\to \pi^0\pi^0) &=&{1\over\sqrt{2}}\lambda_d(-C+E+P_d+\PE_d+\PA_d)+{1\over\sqrt{2}}\lambda_s (P_s+\PE_s+\PA_s), \non \\
A(D^+\to\pi^+\pi^0) &=& {1\over\sqrt{2}}\lambda_d(T+C)
\en
with the decay amplitude of $D^0\to \pi^+\pi^-$ given in Eq.~(\ref{eq:Amp:D0pipi}), we obtain the isospin amplitudes
\be
A_0 &=& {1\over\sqrt{3}}[\lambda_d(2T-C+3E)+\sum_{p=d,s}3\lambda_p(P_p+\PE_p+\PA_p)], \non \\
A_2 &=& \sqrt{2\over 3}\lambda_d(T+C).
\en
Experimentally, the ratio $|A_0/A_2|$ is determined from the relation
\be
{\Gamma(D^+\to \pi^+\pi^0)\over \Gamma(D^0\to\pi^+\pi^-)+\Gamma(D^0\to \pi^0\pi^0)}={3\over 2}\,{|A_2|^2\over |A_0|^2+|A_2|^2}\, .
\en
It follows that $|A_0/A_2|$ is equal to $2.5\pm0.1$ in charm decays and $22.4\pm0.1$ in the kaon system.  To a good approximation,
\be
\left| {A_0\over A_2} \right| \approx {1\over\sqrt{2}}\,{2T-C+3E+3\Delta P\over T+C}.
\en
Neglecting the penguin contribution for the moment and assuming SU(3) symmetry for the topological amplitudes so that we can apply Eq.~(\ref{eq:PP1}) to obtain $|A_0/A_2|=3.8$\,. When SU(3) symmetry breaking in $T$ and $C$ amplitudes \cite{ChengCV2} are taken into account, one would have
\be
\left| {A_0\over A_2} \right| \approx {1\over\sqrt{2}}\,{1.92T-0.79C+3E+3\Delta P\over 0.96T+0.79C}
\en
and the ratio will be reduced to 3.2\,. At any rate, it is evident that the $\Delta I=1/2$ enhancement over the $\Delta I=3/2$ one in $D$ decays arises predominately from the tree amplitudes. As the predicted ratio tends to be too large compared to experiment, this means that the difference $\Delta P$ in penguin contractions should contribute destructively to $|A_0/A_2|$. This is very different from the kaon case where the predicted ratio due to tree amplitudes is too small compared to experiment, and it is quite obvious that a large enhancement of the penguin matrix element is needed in order to explain the discrepancy between theory and experiment for the ratio $|A_0/A_2|$ in $K\to \pi\pi$ decays.

Under the assumption of large enhancement of the penguin amplitude ${1\over 2}\Sigma P$ [see Eq.~(\ref{eq:sum&difP})] relative to the tree one by a factor of $1/\epsilon'$, Brod {\it et al.} argued that $\Sigma P$ could explain $\Delta a_{CP}^{\rm dir}$, while the difference $\Delta P$ explained the large disparity in the rates of $D^0\to K^+K^-$ and $D^0\to\pi^+\pi^-$  \cite{Brod}. This would require that $\Delta P$ be of the same order as the tree amplitude.
A consistent picture for large penguins in SCS $D\to PP$ decays is emerged provided that the relevant amplitudes scale as
\be \label{eq:scaling}
T\sim {\cal O}(1), \qquad {1\over 2}\Sigma P\sim {\cal O}(1/\epsilon'),\qquad \Delta P\sim {\cal O}(\epsilon_U/\epsilon')\sim {\cal O}(1) \ ,
\en
with $\epsilon_U$ being a $U$-spin breaking parameter. While this scenario sounds appealing, we notice that the realistic symmetry breaking effects in amplitudes $T$ and $E$ does not follow the pattern depicted in Eq.~(\ref{eq:UbreakinT}). For example, the factorization approach leads to the robust result
$|T_{_{K\!K}}/T_{\pi\pi}| \approx 1.32$\,. Assuming SU(3) symmetry for $E$ amplitudes, we will have $|\Delta P/T|\sim {\cal O}(0.15)$ rather than $|\Delta P/T|\sim {\cal O}(1)$. And it will become smaller when symmetry breaking in $E$ is taken into account. In realistic model calculations, $|\Delta P/T|$ is very small and negligible. The quantity $\Delta P$ comes from the difference of $d$- and $s$-loop contractions induced mainly from the 4-quark operator $O_1$. Penguin contraction amplitudes read
\be \label{eq:P46}
{\cal P}^p_{4,6}&=&{C_F\alpha_s\over 4\pi N_c}\Bigg\{ c_1\left[ {4\over 3}{\rm ln}{m_c\over \mu}+{2\over 3}-G_{M_2}(r_p,k^2)\right]+\cdots \Bigg\} \ ,  \non
\en
where $p=d,s$, $r_i\equiv m_i^2/m_c^2$, $k^2$ is the momentum squared carried by the virtual gluon and the perturbative loop function $G(r,k^2)$ is given by
\be
G(r,k^2)=-4\int_0^1 du u(1-u)\ln\left[r-u(1-u){k^2\over m_c^2}\right].
\en
Hence,
\be
P_{4,6}^d-P_{4,6}^s={\cal P}_{4,6}^d-{\cal P}_{4,6}^s={C_F\alpha_s\over 4\pi N_c}[G(m_s^2/m_c^2,k^2)-G(m_d^2/m_c^2,k^2)].
\en
By varying $k^2$ between $m_c^2$ and $m_c^2/4$, one can see that $\Delta P$ is rather small, as shown in Eq.~(\ref{eq:Pover T}).

It is often stated in the literature that due to the large $1/m_c$ corrections, it is reasonable to assume an enhancement of the hadronic matrix elements by a factor of, say,  $3\sim 5$. However, we notice that this enhancement is not applicable to the color-allowed tree amplitude $T$ as its prediction based on the factorization approach agrees with experiment even before enhancement. For the amplitude $E$, its large $1/m_c$ power corrections arise from final-state rescattering. Hence, it is the decay amplitude rather than the hadronic matrix element that gets enhanced by $1/m_c$ effects because the long-distance FSI cannot be expressed as a single hadronic matrix element of local 4-quark operators.

We digress here to make a side remark on the so-called $\Delta A_{K\pi}$ puzzle related to the difference of \CP asymmetries of $B^-\to K^-\pi^0$ and $\bar B^0\to K^-\pi^+$. The decay amplitudes of $\bar B\to \bar K\pi$ in terms of topological diagrams read
\begin{eqnarray} \label{eq:ampBKpi}
 A(\bar B^0\to K^-\pi^+) &=& P+T+{2\over 3}P^c_{\rm EW}+P_A, \nonumber \\
 A(B^-\to K^-\pi^0) &=& {1\over\sqrt{2}}(P+T+C+P_{\rm
 EW}+{2\over 3}P^c_{\rm EW}+A+P_A).
\end{eqnarray}
We notice that if $C$, $P_{\rm EW}$ and $A$ are negligible compared with $T$, it is clear from Eq.~(\ref{eq:ampBKpi}) that the decay amplitudes of $K^-\pi^0$ and
$K^-\pi^+$ will be the same, apart from a trivial factor of $1/\sqrt{2}$. Hence,
one will expect that $A_{CP}(K^-\pi^0)\approx A_{CP}(K^-\pi^+)$, while they
differ by 5.6$\sigma$ experimentally,
$\Delta A_{K\pi}\equiv A_{CP}(K^-\pi^0)-A_{CP}(K^-\pi^+)=0.124\pm0.022$ \cite{HFAG}. A large penguin $P$ cannot explain the $\Delta A_{K\pi}$ puzzle as it contributes equally to both $B^-\to K^-\pi^0$ and $\bar B^0\to K^-\pi^+$. This puzzle will be resolved provided that $c/T$ is of order
$1.3\sim 1.4$ with a large negative phase, where $c\equiv C+P_{\rm EM}$
($|c/T|\sim 0.9$ in the standard short-distance effective Hamiltonian approach) \cite{Cheng2009}. There exist two popular scenarios for achieving a
large complex $c$: either a large complex $C$ or a large complex $P_{\rm EW}$ (see \cite{Cheng2009} for references and details). However, for a large complex $P_{\rm EW}$ one needs NP beyond the SM because $P_{\rm EW}$ is essentially real in the SM as it does not carry a nontrivial strong phase. In principle, one cannot discriminate between these two scenarios in penguin-dominated decays as $C$ and $P_{\rm EW}$ are comparable in magnitude due to the CKM enhancement for the latter. Nevertheless, these two scenarios will lead to very distinct predictions for tree-dominated decays where $P_{\rm EW}\ll C$. For example, the decay rates of $\bar B^0\to \pi^0\pi^0,\rho^0\pi^0$ will be substantially enhanced for a large $C$ but remain intact for a large $P_{\rm EW}$.
Owing to the large branching fractions observed for $\bar B^0\to \pi^0\pi^0,\rho^0\pi^0$, it is more appealing to resolve the $B\to K\pi$ \CP puzzle using a large complex $C$ instead of invoking NP on electroweak penguins.

\section{New Physics Effects \label{sec:NP}}

If the experimental measurement continues to be large with more statistics in the future, it will imply new physics beyond the SM in the charm sector. Then it will be important to explore possible NP scenarios responsible for such large direct \CP asymmetries.

The surprising LHCb measurement has inspired many different analyses based on a variety of NP models. Scenarios for NP effects at tree level include flavor-changing coupling of a SM $Z$ boson \cite{Giudice,Altmannshofer}, flavor-changing neutral currents induced by a leptophobic massive $Z'$ boson \cite{Zhu,Altmannshofer}, two Higgs-doublet model \cite{Altmannshofer}, color-singlet scalar model \cite{Nir}, color-sextet scalar model \cite{Altmannshofer,Chen}, color-octet scalar model \cite{Altmannshofer} and fourth generation model \cite{Rozanov,Feldmann}. Models with NP in QCD penguins at the loop level have been constructed as well, including new fermion and scalar fields \cite{Altmannshofer} and the chirally enhanced chromomagnetic dipole operator \cite{Giudice}. \CP violation induced by supersymmetric $R$-parity violating interactions has also been considered in \cite{LiuC}.

It is known that NP models are highly constrained by $D^0$-$\ov D^0$ mixing, $K^0$-$\ov K^0$ mixing and \CP violation in the kaon system characterized by the parameter $\epsilon'/\epsilon$ \cite{Ligeti}. Many of the tree-level NP models are either ruled out or in tension with flavor-related experiments \cite{Altmannshofer}. As pointed out in \cite{Giudice}, a large NP contribution to the $\Delta C=1$ chromomagnetic dipole operator is probably the best candidate to explain the LHCb and CDF results as it is least constrained by all current data in flavor physics.

In \cite{ChengCV2} we have considered two possibilities of new physics effects, namely, large penguins and large chromomagnetic dipole operator, and studied their phenomenological consequences in the \CP asymmetries of SCS $D\to hh$ decays, seeing if there are discernible differences in the two scenarios.

\subsection{Large penguins}
It is known that a large penguin of order $3T$ can explain the observed $\Delta a_{CP}^{\rm dir}$ \cite{Brod,Bhattacharya:2012}. More precisely, the penguin amplitudes
\be \label{eq:largeP}
{1\over 2}\Sigma P\approx
\left\{
\begin{array}{ll}
2.9\,T e^{i 85^\circ} & \mbox{for Solution I}  \\
3.2\,T e^{i 85^\circ} & \mbox{for Solution II}
\end{array}
\right.
\en
with nearly maximal strong phase can accommodate the measurement of direct \CP asymmetry difference between $D^0\to K^+K^-$ and $\pi^+\pi^-$ (see Table~\ref{tab:CPVpp_new}). This can be easily seen from the approximated relation
\be
\Delta a_{CP}^{\rm dir}\approx -2.4\times 10^{-4}\left|{{1\over 2}\Sigma P\over T+E}\right|\sin\delta,
\en
derived from Eq.~(\ref{eq:pipiacp}), where we have applied the approximation of $a_{CP}^{\rm dir}(\pi^+\pi^-)\approx -a_{CP}^{\rm dir}(K^+K^-)$.
Using the above large penguin ${1\over 2}\Sigma P$ as input, the predicted direct \CP asymmetries for other SCS charm decays are summarized in the second column of Table~\ref{tab:CPVpp_new}. We see that many modes, such as $D_s^+\to \pi^+K^0,\pi^0 K^+,K^+\eta'$, are expected to yield direct \CP asymmetries of a similar magnitude, at a few per mille level.

Penguin amplitudes could be enhanced due to some nonperturbative effects in the SM \footnote{Authors of \cite{Bhattacharya:2012} have introduced an additional phenomenological penguin amplitude $P_b$ in order to accommodate the measured $\Delta a_{CP}^{\rm dir}$. However, there is no need to make this assumption as the penguin amplitude can be recast to
$\lambda_d P_d+\lambda_s P_s={1\over 2}(\lambda_d-\lambda_s)(P_d-P_s)-{1\over 2}\lambda_b(P_d+P_s)$. The second term on the RHS of the above relation is the so-called $P_b$ in \cite{Bhattacharya:2012}.
}
or NP effects beyond the SM. The predictions in Table~\ref{tab:CPVpp_new} are presented irrespective of the origin of enhancement. However, whether or not the large penguins are subject to the constraints from $D^0$-$\ov D^0$ mixing, $K^0$-$\ov K^0$ mixing, etc. remains to be investigated.

\subsection{Large chromomagnetic dipole operator}
The authors in \cite{Giudice} have argued that a large chromomagnetic dipole operator could be the best NP candidate to explain the data while satisfying most flavor physics constraints at the same time.
Although the chromomagnetic dipole operator $O_{8g}$ is suppressed by the charm Yukawa coupling, the hadronic matrix element
$\la M_1M_2|O_{8g}|D\ra$ scales as $m_c^2/k^2$,
where $k^2$ is the square of momentum transfer of the gluon and is of order $m_c^2$. As a consequence, the matrix element is independent of $m_c$; that is, it is enhanced by a factor of $1/m_c$. After absorbing the factor $m_c$ into the definition of $O_{8g}$ from new physics loop contribution, such as a low-energy supersymmetry scenario discussed in detail  in \cite{Giudice}, the corresponding Wilson coefficient looks like being enhanced by a factor of $v/m_c$, where $v$ is the vacuum expectation value of the Higgs field and used to represent the typical new physics scale.
On the contrary, the $D^0$-$\ov D^0$ mixing induced by $O_{8g}$ is suppressed by a factor of $m_c^2/v^2$. This illustrates why the dipole operator can escape the constraint from $D^0$-$\ov D^0$ mixing.
In short, we need NP to enhance the Wilson coefficient $c_{8g}$ and to induce a sizable imaginary part. This can be realized in the supersymmetric models where the gluino-squark loop contributes a major part of $c_{8g}$ \cite{Kagan}, the disoriented $A$ terms and split families are the sources of flavor violation \cite{Giudice}, or the flavor structure of the trilinear scalar couplings is related to the structure of the Yukawa couplings via approximate flavor symmetries \cite{Hiller}, or the supersymmetric realization of partial compositeness \cite{Boaz}.

For the purpose of illustration, we shall take $c_{8g}^{\rm NP}\approx 0.012e^{i14^\circ}$ which fits to the data of $\Delta a_{CP}^{\rm dir}$. The calculated \CP asymmetries for the other modes are listed in the last column of Table~\ref{tab:CPVpp_new}.
It is interesting to notice that while a large chromomagnetic dipole operator leads to large direct \CP asymmetry for $D^0\to\pi^0\pi^0,\pi^0\eta$, the predicted \CP violation for $D^0\to \pi^0\eta'$, $D^+\to\pi^+\eta', K^+\ov K^0$ and $D_s^+\to \pi^+K^0,K^+\eta'$ is much smaller than that in the large penguin scenario. Therefore, measurements of the \CP asymmetries of the above-mentioned modes will enable us to discriminate between the two different NP scenarios.


\begin{table}[t]
\begin{center}
\begin{tabular}{ l   c c} \hline\hline
Decay Mode~~ & Large penguins~~~ & Large c.d.o. \\
 \hline
$D^0\to \pi^+ \pi^-$ & 3.96 (4.40)~~~ & 5.18 (3.70) \\
$D^0\to\pi^0 \pi^0$   & 0.93 (1.01)~~~ & 8.63 (6.19)  \\
$D^0\to \pi^0 \eta $  & 0.09 (0.03)~~~ & $-6.12$ ($-4.15$) \\
$D^0\to \pi^0 \eta'$  & 2.36 (2.67)~~~ & $-0.44$ ($-0.44$) \\
$D^0\to \eta\eta $ &  $-1.79$ ($-1.64$)~~~ & $-1.63$ ($-2.00$)\\
$D^0\to \eta\eta' $  & $2.65$ (1.49)~~~  & $-2.30$ ($-1.08$) \\
$D^0\to K^+ K^{-}$   & $-2.63$ ($-2.36$)~~~ & $-1.46$ $(-2.88)$ \\
$D^+\to \pi^+ \pi^0$   & 0 (0)~~~ & 0 (0) \\
$D^+\to \pi^+ \eta $  &  $-3.24$ $(-3.62)$~~~ & $-5.35$ $(-3.67)$   \\
$D^+\to\pi^+ \eta' $  & $2.97$ (3.34)~~~ &  0.93 (0.59) \\
$D^+\to K^+ \ol{K}^{0}$  &  $-2.95$ ($-3.28$)~~~ & 0.37 (0.29)  \\
$D_s^+\to \pi^+ K^{0}$  & $3.29$ (3.66)~~~ & $-0.47$ $(-0.35)$  \\
$D_s^+\to \pi^0 K^{+}$  &  $4.57$ (5.08)~~~ & 4.40 (3.14)\\
$D_s^+\to K^{+}\eta$   &  $-0.58$ ($-0.57$)~~~ & 1.59 (0.94) \\
$D_s^+\to K^{+}\eta' $  &  $-5.16$ ($-5.79$)~~~ & 1.76 (1.39)\\
\hline\hline
\end{tabular}
\caption{\small Direct \CP asymmetries (in units of $10^{-3}$) of SCS $D\to PP$ decays estimated in the scenarios with large penguin contributions and large chromomagnetic dipole operator.  The parameters $\Sigma P$ and $c_{8g}^{\rm NP}$ are chosen to fit the data of $\Delta a_{CP}^{\rm dir}$:
${1\over 2}\Sigma P=2.9\,Te^{i85^\circ}$ and $c_{8g}^{\rm NP}=0.017e^{i14^\circ}$ for Solution I,  ${1\over 2}\Sigma P=3.2\,Te^{i85^\circ}$ and  $c_{8g}^{\rm NP}=0.012e^{i14^\circ}$ for Solution II.
  \label{tab:CPVpp_new}}
\end{center}
\end{table}

\subsection{Discrimination between SM and NP interpretations}

In view of two possible explanations of large \CP asymmetry difference $\Delta a_{CP}^{\rm dir}$ by the SM and by NP, it is natural to ask if there are experimental tests that allow us to distinguish between the SM and NP interpretations. There are a few proposals for this purpose: (i) Suppose NP has $\Delta I=3/2$ contributions. If electroweak penguin and isospin breaking effects are negligible, then a measurement of nonzero $a_{CP}^{\rm dir}$ in $D^+\to \pi^+\pi^0$ will be a signal for $\Delta I=3/2$ new physics \cite{Zupan}. Some sets of isospin sum rules can be constructed along this direction. (ii) If a large chromomagnetic dipole operator is responsible for $\Delta a_{CP}^{\rm dir}$, the corresponding electromagnetic dipole operator will also get enhanced. Direct \CP asymmetries in radiative $D\to P^+P^-\gamma$ decays with the invariant mass $M_{P\!P}$ close to the $\rho$ or the $\phi$ peak can be achieved at the level of several percent \cite{Isidori}. Hence, it was claimed in \cite{Isidori} that evidence of $|a_{P\!P\gamma}|\geq 3\%$ would be a clear signal of physics beyond the SM and a clear indication of new \CP-violating dynamics associated to dipole operators.

Since the SM interpretation of large $\Delta a_{CP}^{\rm dir}$ relies on the large penguin amplitude which is about 3 times bigger than the color-allowed tree amplitude, the predictions shown in Table~\ref{tab:CPVpp_new} will also serve the purpose of discriminating the SM from the NP model based on the chromomagnetic dipole operator. As elaborated on in the previous subsection,
SM and NP in dipole operators can be distinguished as their predictions for \CP asymmetries in $D^0\to\pi^0\pi^0,\pi^0\eta,\cdots, D_s^+\to K^+\eta'$ appear to be ``orthogonal" to each other.

\subsection{Other decay modes of interest}

Since the large direct \CP asymmetry difference in $D^0\to K^+K^-$ and $D^0\to \pi^+\pi^-$ has been observed, one may ask if such an effect also occurs in other decay modes. From Table~\ref{tab:CPVpp_new} and the measured branching fractions of SCS $D\to PP$ decays \cite{PDG}, it turns out that $D^+\to K^+\ov K^0$ and $D_s^+\to \pi^+K^0$ will be good targets for this purpose. However, direct \CP violation in these charged $D$ decays will be contaminated by $K^0$-$\ov K^0$ mixing due to the neutral kaon $K_S$ in the final state.
The decays $D^0\to\pi^0\pi^0$, $D^+\to\pi^+\eta^{(')}$ and $D_s^+\to K^+\eta^{(')}$ will be the next of interest. Observation of \CP violation in the decay $D^0\to K^0\ov K^0$ will be quite interesting, but its rate is just too small.

As for SCS $D\to VP$ decays, one should measure \CP violation in $D^0\to \pi^+\rho^-,\pi^-\rho^+$, $K^+K^{*-},K^-K^{*+}$ to see if their \CP asymmetry differences are comparable to that measured in $D^0\to K^+K^-$ and $D^0\to\pi^+\pi^-$. It is worth mentioning that $\Delta A_{CP}$ is measured to be $(0.51\pm0.28\pm0.05)\%$ in $D^+\to \phi\pi^+$ by Belle \cite{Belle}. Naively, its direct \CP violation is expected to vanish as $D^+\to \phi\pi^+$ proceeds only through the color-suppressed tree diagram $C_P$ where the subscript $P$ indicates that the pseudoscalar meson contains the spectator quark of the charmed meson. Nevertheless, this decay also receives singlet QCD-penguin contribution $S$ [see Fig.~\ref{Fig:Quarkdiagrams}(d)] governed by the effective parameters $a_3$ and $a_5$. The interference between $C_P$ and $S$ will allow the presence of \CP asymmetry in this mode.

From the experimental point of view, it is important to search for \CP-violating effects in three- and four-body charm decays. For example, the Dalitz-plot analysis of $D^+\to K^+K^-\pi^+$ allows us to differentiate strong phases from the weak ones across the Dalitz plot.
However, theoretically it will be difficult to estimate these effects due to the lack of information on $1/m_c$ power corrections. The model-independent diagrammatic approach has not been proved to be applicable to three-body decays.

\section{Conclusions}

Based on the topological diagram approach for tree amplitudes and QCD factorization for a crude estimation of perturbative penguin amplitudes, we have studied direct \CP asymmetries in singly Cabibbo-suppressed $D$ decays within the framework of the SM and concluded that the \CP asymmetry difference $\Delta a_{CP}^{\rm dir}$ between $D^0 \to K^+ K^-$ and $D^0 \to \pi^+ \pi^-$ is of order $-(0.14\sim 0.15)\%$. We have explored the phenomenological implications of two new physics scenarios for explaining the observed \CP asymmetry in the charm sector, one with large penguin amplitudes and the other with a large chromomagnetic dipole operator.

\Acknowledgements
I am grateful to the organizers for this wonderful and successful conference and to Cheng-Wei Chiang for fruitful collaborations and for reading the manuscript.

\end{document}